\documentclass[aps,pra,twocolumn,groupedaddress,floatfix]{revtex4-1}

\usepackage{graphicx}
\usepackage{dcolumn}
\usepackage{bm}
\usepackage{amsmath}
\usepackage{color}

\begin{document}

\title{Trial wave functions for ring-trapped ions and neutral atoms: 
Microscopic description of the quantum space-time crystal}

\author{Constantine Yannouleas}
\email{Constantine.Yannouleas@physics.gatech.edu}
\author{Uzi Landman}
\email{Uzi.Landman@physics.gatech.edu}

\affiliation{School of Physics, Georgia Institute of Technology,
             Atlanta, Georgia 30332-0430}

\date{21 September 2017}

\begin{abstract}
A constructive theoretical platform for the description of quantum space-time crystals uncovers for $N$ 
interacting and ring-confined rotating particles the existence of low-lying states with proper space-time crystal 
behavior. The construction of the corresponding many-body trial wave functions proceeds first via symmetry breaking at
the mean-field level followed by symmetry restoration using projection techniques. 
The ensuing correlated many-body wave functions 
are stationary states and preserve the rotational symmetries, and at the same time they reflect the point-group 
symmetries of the mean-field crystals. This behavior results in the emergence of sequences of select magic angular 
momenta $L_m$. For angular-momenta away from the magic values, the trial functions vanish. Symmetry breaking beyond 
mean field can be induced by superpositions of such good-$L_m$ many-body stationary states. We show that 
superposing a pair of adjacent magic angular momenta states leads to formation of special broken-symmetry states 
exhibiting quantum space-time-crystal behavior. In particular, the corresponding particle densities rotate around the 
ring, showing undamped and nondispersed periodic crystalline evolution in both space and time. The experimental 
synthesis of such quantum space-time-crystal wave packets is predicted to be favored in the vicinity  of ground-state 
energy crossings of the Aharonov-Bohm-type spectra accessed via an externally applied magnetic field. These results are
illustrated here for Coulomb-repelling fermionic ions and for a lump of contact-interaction attracting bosons.
\end{abstract}

\maketitle

\section{Introduction}

Groundbreaking experimental progress 
\cite{bloc08,blat12,spie14,taba15,haef15,haef17,nogu14,bosh14,moul12,phil11,note1} 
in the field of trapped ultracold ions and neutral atoms, in 
particular the unprecedented control of interpaticle interactions and the attainment of ultracold temperatures, 
offer these systems as prime resources for experimental realization of the emergent exciting concept of a  quantum 
space-time crystal (QSTC). Inspired by the relativistic 3+1-dimensions analogy \cite{wilc12}, 
\textcolor{black}{
the QSTC idea extends translational symmetry breaking (SB) to encompass both the spatial and time dimensions.}
Indeed, the original QSTC proposal 
\cite{wilc12,li12} motivated an abundance of scientific discussion, commentary, and exploration
\cite{brun13,brun13.2,nozi13,wilc13,li13,wata15,sach15,sond16,naya16,yao16,fazi17}. 

The original QSTC was proposed in the form of crystalline spatial-particle-density arrangements \cite{li12}, or 
other solitonic-type (charge-density-wave) formations \cite{wilc12} revolving around a ring-shaped ultracold trap 
without dispersion or damping. 
Although significant experimental progress has been reported toward this goal \cite{haef15,haef17}, formation of a 
QSTC in this experimental configuration is yet to be demonstrated. At the same time, experimental progress for a 
``weaker class'' \cite{gibn17} of discrete-time-crystals \cite{sach15,sond16,naya16,yao16,fazi17} limited 
exclusively to the time domain has been reported \cite{monr17,luki17}, employing time-periodically-driven spin 
systems. Contributing to this state of affairs are 
\textcolor{black}{limitations} 
of earlier theoretical treatments of the QSTC 
that were discussed extensively in previous commentary \cite{brun13,brun13.2,nozi13,wata15}, e.g., limiting oneself 
to mean-field (MF) dynamics \cite{wilc12}, or considering solely the energetics of states with good total angular 
momenta which (as a matter of principle) have uniform spatial densities \cite{li12,li12.2}. To throw further light 
on the nature and properties of QSTCs, it is imperative that a formulation and implementation of appropriate 
many-body trial wave functions for the QSTC on a ring be advanced. The sought-after trial wave functions should 
explore for a finite system of $N$ particles the interplay \cite{yann07} between the mean-field symmetry-broken 
states, which are not eigenstates of the total angular momentum $\hat{L}$, and the exact symmetry-preserving (good 
total-angular-momentum) states.

Here, we introduce such trial wave functions and analyze their spectra and combined spatially dispersionless and 
temporally undamped evolution,
\textcolor{black}{which are the defining characteristics of a QSTC.}
Contrasting with these findings, 
previous beyond-mean-field theoretical studies \cite{degu12,kavo17,cede11,carr16,schm17} that investigated spatial 
solitonic formations in finite boson systems in one dimension or on a ring have revealed drastically different  
behaviors, such as increasing dispersion with time accompanied by a revival at the initial position of the propagated
inhomogeneous wave packet \cite{robi15}.  

We employ a beyond-mean-field methodology of symmetry restoration via projection techniques, introduced by us 
previously \cite{yann02,yann02.2,yann03.2,yann03,yann04,yann04.2,yann06,yann07,yann09,yann11} 
for two-dimensional semiconductor quantum dots (with and without an applied magnetic field ${\bf B}$). The 
multilevel symmetry-breaking and symmetry-restoration approach which we persue provides a 
complete theoretical framework for treating symmetry breaking aspects in
finite systems, without reference to the $N \rightarrow \infty$ limit. Indeed this approach originated, and is
widely employed, in nuclear physics and chemistry 
\cite{yann07,peie57,ringbook,lowe55,robl02,bend03,doba07,sun16,ring00,naza12}.

The paper is organized as follows. In Section II we introduce and illustrate the hierarchical, multilevel methodology
that we use for the construction of the trial wave functions for the microscopic many-body Hamiltonian of few  
ultracold ring-confined interacting particles. Following a short synopsis of the method, we discuss first in section 
II A the mean-field, broken-symmetry state, and subsequently in section II B a beyond-mean-field level is outlined,
entailing symmetry-restoration via the use of an angular momentum projection technique.
\textcolor{black}{This results}
in many-body 
stationary-state good-angular-momentum solutions of the microscopic Hamiltonian. These (projected) symmetry-restored 
states show uniform particle density around the ring. However, simultaneously they posses hidden crystalline symmetries 
which can be revealed through the analysis of the corresponding conditional probability densities. In section III C 
we complete our exposition of the construction of the QSTC trial wave functions by analyzing the properties of 
superpositions of pairs of the above-noted symmetry-restored (projected) stationary states [see Eq.\ (\ref{pwm})] that 
are favored to mix in the vicinity of crossings of Aharonov-Bohm-type spectra of ground-state energies versus 
applied magnetic-field (Fig.\ \ref{fspec}). The particle density corresponding to such superposed wavefunctions reveals
crystalline structure on the ring.  Numerical solutions using the trial wave functions are illustrated and analyzed 
for the case of few (even and odd in number) Coulomb-repelling fermionic ions, and for a lump of contact-interacting 
attractive bosons. When evolved with the microscopic many-body Hamiltonian, these trial functions exhibit, for both 
the fermionic repelling ions and attracting bosons, undamped and non-dispersive space and time crystalline periodic 
evolution -- that is, they exhibit breaking of both the space and time symmetries. 

Section III is devoted to further elaboration
\textcolor{black}{on}
three main topics. In section III A we discuss the symmetry 
properties of the symmetry-restored (projected) wave functions and the selection rules for their ``magic'' angular 
momenta. Section III B analyzes the properties of the initial wave packets and their associated time evolution, and 
section III C comments on the relation between the constructed trial functions (in particular  the aforementioned 
symmetry-restored stationary states) and the wavefunctions obtained through exact-diagonalization 
[configuration-interaction (CI)] solutions of the microscopic many-body Hamiltonian. 

We conclude in section IV with a summary of our work, including a brief listing of recent progress achieved in 
developing experimental techniques for preparation and measurement of ring-confined ultracold particles. The 
Appendices give tables of numerical results (rotational energies for different magic angular momenta, and moments of 
inertia) for the systems investigated in the paper, as well as explicit expressions for the conditional probability 
distribution and single particle density. 

\section{The interplay between symmetry-broken and symmetry-preserving states: Group-theoretical formulation}
\label{sbsp}

In connection with the QSTC, we consider three levels of many-body trial wave functions: (1) A Slater determinant for 
localized fermions (or permanent for localized bosons) on the ring. We denote this wave function by $\Psi^{\rm SB}$; it 
corresponds to the unrestricted Hartree-Fock (UHF), or Gross-Pitaevskii (GP), mean-field step \cite{yann07}
that exhibits symmetry breaking of the space degrees of freedom. $\Psi^{\rm SB}$ does not preserve the total
angular momentum. Out of the three levels in the hierarchical scheme (see below), it is the trial wave function 
closest to the familiar concept of a classical Wigner crystal \cite{giulbook}. 
(2) A stationary multideterminantal (multipermanental) 
wave function $\Phi^{\rm PROJ}_L$ characterized by a good total angular momentum $\hbar L$, which is generated 
by applying a projection operator ${\cal P}_L$ (see below) on 
$\Psi^{\rm SB}$. This step goes beyond the MF approximation and restores (as required) the quantum many-body
Hamiltonian symmetries in the stationary-state solutions. Unlike $\Psi^{\rm SB}$, $\Phi^{\rm PROJ}_L$ exhibits an 
azimuthally uniform single-particle density [SPD, $\rho({\bf r},t)$], which is also time-independent (stationary).
Previously, we referred to such projected wave functions $\Phi^{\rm PROJ}_L$ as quantum rotating Wigner 
molecules \cite{yann09}. 
\textcolor{black}{
(3) Coupling between the stationary states (brought about by a perturbation which we term in the following as 
``the pinning agent'') results in a superposition of two projected wave functions with different angular momenta 
$L_1$ and $L_2$, leading to formation of a pinned Wigner molecule (PWM), i.e.,} 
\begin{equation}
\Phi^{\rm PWM}(L_1,L_2; t=0) = 
\alpha \Phi^{\rm PROJ}_{L_1} + \beta e^{i\phi(t=0)}\Phi^{\rm PROJ}_{L_2},
\label{pwm}
\end{equation} 
where $\phi(t=0)$ can be set to zero without loss of generality, and $\alpha^2+\beta^2=1$. 
\textcolor{black}{ 
In the following we illustrate the case of $\alpha=\beta =1/\sqrt{2}$ (the physics of the PWM maintains for other 
choices of the mixing coefficients).} 
For selected magic (see below) $L_1$ and $L_2$, $\Phi^{\rm PWM}(L_1,L_2; t=0)$ represent a special family of quantal 
wave packets with broken azimuthal symmetry. 
Consequently their corresponding $\rho({\bf r},t=0)$ are not uniform, forming 
instead a crystal-like particle density pattern, with $kN$, $k=1,2,3,\ldots$ possible peaks for $N$ fermionic ions 
and $1,2,3,\ldots$ possible peaks for $N$ attractive bosons. When the pinning agent is lifted, the 
$\Phi^{\rm PWM}(L_1,L_2;t)$ evolve in time undamped according to the exact many-body quantum Hamiltonian dynamics, 
i.e., the phase $\phi$ will vary as $\phi(t)=(E_2-E_1) t/\hbar$, and the associated $\rho({\bf r},t)$ will 
oscillate at any given space point with a time period $T=\tau/n$; $\tau=2\pi\hbar/|E_1-E_2|$, $E_{1(2)}$ 
being the energies of the stationary states $\Phi^{\rm PROJ}_{L_i}$ with $i=1,2$, respectively. 
For either statistics (fermions or bosons), $n = N$ for repelling ions and $n = 1$ for attractive particles; 
as aformentioned, here we discuss explicitly Coulomb-repelling fermionic ions and contact-interacting attractive 
bosons. Such undamped and dispersionless periodic time variation is not possible for the MF Hartree-Fock (or 
Gross-Pitaevskii) wave packet $\Psi^{\rm SB}$, because it contains all the possible angular momenta when expanded 
in the complete basis set of the stationary wave functions $\Phi^{\rm PROJ}_L$. Additionally, the MF wave functions 
lose \cite{grif76} their single-determinant (single-permanent) character under the exact time evolution. 

The many-body Hamiltonian of $N$ identical particles in a ring-type trap threaded by a constant 
magnetic field ${\bf B}$ is
\begin{align}
\begin{split}
{\cal H}= \sum_{i=1}^N & \biggl( \frac{({\bf p}_i-\eta {\bf A}_i)^2}{2M} +
\frac{(r_i-R)^2}{2l_0^2/(\hbar \omega_0)} \biggr) + \sum_{i < j} V( r_{ij}),
\end{split}
\label{hmb}
\end{align}
where ${\bf A}({\bf r})={\bf B}{\times}{\bf r}/2$ is the vector potential in the symmetric gauge,
$r=\sqrt{x^2+y^2}$, $\omega_0$ is the frequency of the trap, $R=\sqrt{X^2+Y^2}$ is the ring radius, 
the oscillator length $l_0=\sqrt{\hbar/(M\omega_0)}$, and $r_{ij}=|{\rm r}_i-{\rm r}_j|$. ${\bf B}$ can be the 
familiar magnetic field in the case of charged ions (when $\eta=e/c$), or a synthetic one in the case
of ultracold neutral atoms \cite{spie14}.  

\textcolor{black}{
\subsection{First level: The mean-field-ansatz, symmetry-broken crystalline state}}
\label{fl}

\textcolor{black}{
In view of the strong inter-particle interactions (large values of the parameters $R_W$ and $R_\delta$, see below) 
associated with the proposed experimental realizations of the QSTC's \cite{wilc12,li12}, resulting, for ultracold 
quasi-1D-ring-trapped particles, in formation 
of a Wigner crystal (in the case of Coulomb repelling ultracold ions) or a lump (in the case of contact 
attracting bosons), we construct here the symmetry-broken initial state via the use of an ansatz. This ansatz explores 
the localized nature of the $N$ space orbitals (corresponding to the $N$ ring-trapped particles) from which the 
single broken-symmetry determinant (permanent) is formed in the corresponding fermion (boson) systems. 
Such ansatz proved most adequate to approximate broken-symmetry mean-field solutions (reflecting individual
particle localization) in previous studies of fermions and bosons in harmonically confined quantum dots and other 2D 
systems \cite{yann07,maki83}.}
 
We describe each particle localized at position ${\bf R}_j$ as a displaced Gaussian function
\begin{align}
u({\bf r}, {\bf R}_j) =\frac{1}{\sqrt{\pi} \lambda} \exp \biggl( -\frac{({\bf r}-{\bf R}_j)^2}{2\lambda^2}
-i \varphi({\bf r},{\bf R}_j; B) \biggr),
\label{uorb}
\end{align}
with $\lambda=\sqrt{\hbar/(M\Omega)}$; $\Omega=\sqrt{\omega_0^2+\omega_c^2/4}$ where 
$\omega_c=\eta B/M$ is the cyclotron frequency. The phase in Eq.\ (\ref{uorb}) is due to the gauge 
invariance of magnetic translations \cite{landbook,peie33}) and is given by $\varphi({\bf r},{\bf R}_j; B)=
(xY_j-yX_j)/(2l_B^2)$, with $l_B=\sqrt{\hbar /(\eta B)}$ being the magnetic length. For simplicity, in the
following we provide examples for only three cases: (i) that of $N$ fully polarized fermionic ions with odd 
$N$, (ii) that of $N$ fully polarized fermionic ions with even $N$, and (iii) that of $N$ spinless bosons interacting 
via an attractive contact potential. As will be shown explicitly, case (ii) presents different characteristics compared 
to case (i).

In the case of ultracold ions repelling each other via the Coulomb interaction, we take the 
${\bf R}_j=R_{\rm eq} e^{2\pi (j-1) i/N}$, $j=1,2,\ldots,N$ to coincide 
with the equilibrium positions (forming a regular polygon) of $N$ classical charges inside the annular 
confinement specified in Eq.\ (\ref{hmb}). Then $R_{eq}$ $(> R)$ is given by the real solution of the
cubic equation $aw^3+bw^2+d=0$, where $a=1$, $b=-R$, $d=-l_0^3 R_W S_N/4$, with the Wigner parameter
(the ratio between the characteristic interparticle repulsion and the kinetic zero-point energy of the 
ring-confined particle), $R_W=e^2/(l_0 \hbar \omega_0)$ \cite{yann07} and $S_N=\sum_{j=2}^N 1/\sin[(j-1)\pi/N]$.
Then the corresponding MF wave function, $\Psi^{\rm SB}$, is the determinant formed by the $N$ orbitals 
$u({\bf r}_i, {\bf R}_j)$.

In the case of $N$ ultracold neutral bosons attracting each other with a contact interaction 
$-g \delta({\bf r}_i-{\bf r}_j)$,
\textcolor{black}
{the atoms are localized at the same}
position, and thus $R_{\rm eq}=R$ and
$R_j=R e^{i\theta_0}$, $j=1,2,\ldots,N$. Then the MF wave function, $\Psi^{\rm SB}$, is the product (permanent) of
the orbitals $u({\bf r}_i, R e^{i\theta_0})$. The parameter corresponding to $R_W$ is given here by 
$R_\delta =  g M/\hbar^2$.\\

\subsection{Second level (beyond mean field): The projected, symmetry-restored stationary state}
\label{sl}

A stationary many-body state that preserves the total angular momentum, as well as the rotational symmetry of the 
annular trap, can be projected out of the symmetry-broken $\Psi^{\rm SB}$ by applying the projector operator 
${\cal P}_L$,
\begin{align}
{\cal P}_L = \frac{1}{2\pi} \int_0^{2\pi} e^{i \gamma (L-\hat{L})} d\gamma,
\label{prjop}
\end{align}
where $\hat{L}=\sum_{i=1}^N \hat{l}_i$, $i=1,2,\ldots,N$, and $\hbar \hat{L}$ is the total angular-momentum 
operator. Then the projected many-body state is given by
\begin{align}
\Phi^{\rm PROJ}_L = 
\frac{1}{2\pi} \int_0^{2\pi} d\gamma \Psi^{\rm SB}(\gamma) e^{i\gamma L}.
\label{prj}
\end{align}

${\cal P}_L$ is analogous to the projector operators used in chemistry for molecular orbitals governed by point 
group symmetries \cite{yann03,wolbbook,cottbook,hamebook}. Such projection operators are constructed 
through a summation over the characters of the point group \cite{yann03,cottbook,hamebook}; 
the phases $e^{i\gamma L}$ are the characters of the rotational group in two dimensions \cite{yann03,hamebook} 
and the operator $e^{-i \gamma \hat{L}}$ is the corresponding generator of 2D rotations. Alternatively, Eq.\ 
(\ref{prj}) may be viewed as a linear superposition of all the (energy-degenerate) symmetry-broken states 
$\Psi^{\rm SB}(\gamma)$, azimuthally rotated by $\gamma$. Due to the rotational symmetry, the coefficients of
this superposition, i.e., the phases $e^{i\gamma L}$, can be determined a priori, without the
need to diagonalize a Hamiltonian matrix.    

The projected energies, associated with the stationary wave functions $\Phi^{\rm PROJ}_L$, are given by
\begin{align}
E^{\rm PROJ}(L)=\int_0^{2\pi} h(\gamma) e^{i\gamma L}d\gamma \bigg/ 
\int_0^{2\pi} n(\gamma)e^{i\gamma L} d\gamma,
\label{eprj}
\end{align}
where 
\begin{equation}
h(\gamma)=\langle \Psi^{\rm SB}(0)|{\cal H}| \Psi^{\rm SB}(\gamma) \rangle,
\label{hme}
\end{equation}
and the norm overlap
\begin{equation}
n(\gamma)=\langle \Psi^{\rm SB}(0)| \Psi^{\rm SB}(\gamma) \rangle
\label{nov}
\end{equation}
enforces proper normalization of $\Phi^{\rm PROJ}_L$. Note that the original double integration reduces to a 
single integration over $\gamma$ because ${\cal P}_L^2={\cal P}_L$, $[{\cal P}_L,{\cal H}]=0$.
\textcolor{black}{
We note that the unresticted HF energies for the ansatz determinant (or permanent), $\Psi^{\rm SB}$, before projection
are simply given by
\begin{equation}
E_{\rm UHF}=h(0)/n(0).
\label{euhf}
\end{equation}
}
\begin{figure}[t]
\centering\includegraphics[width=8.0cm]{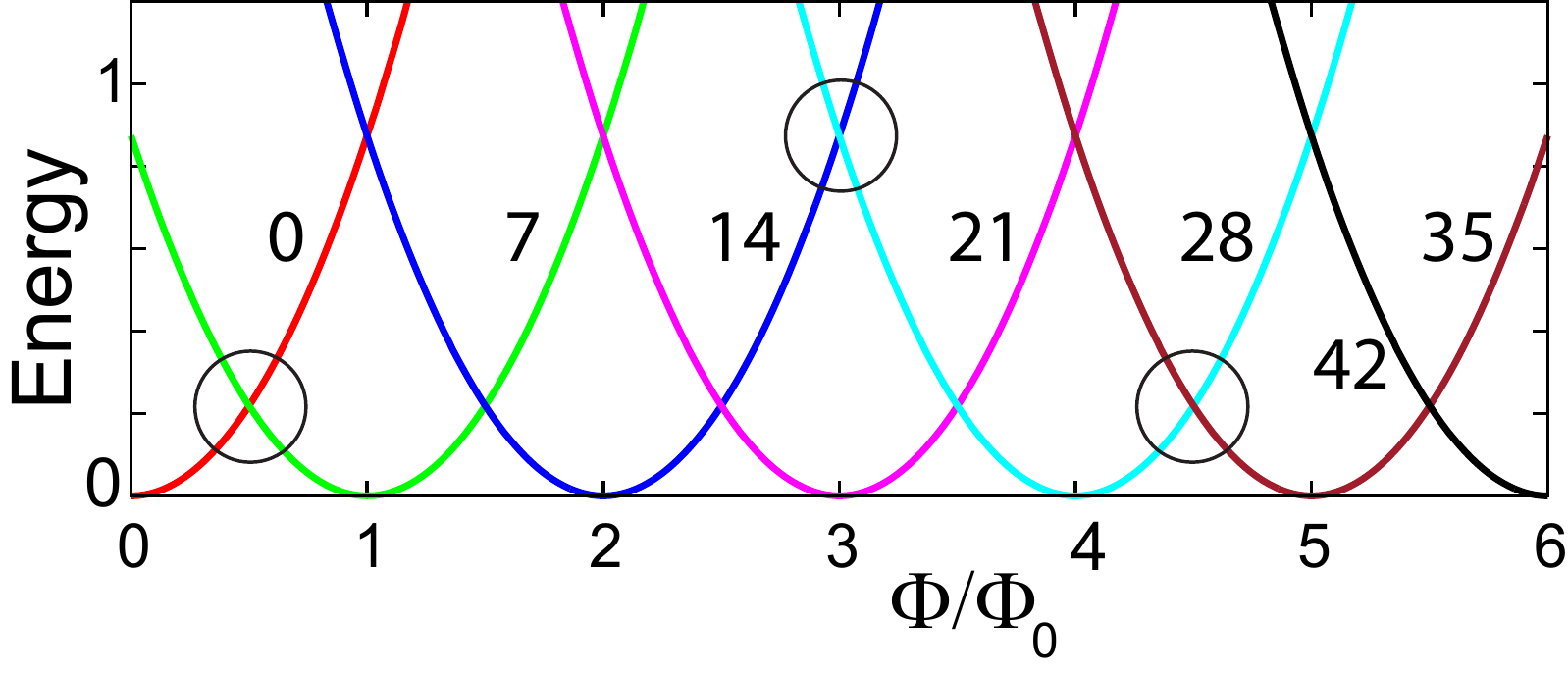}
\caption{
Aharonov-Bohm-type, quantum-rigid-rotor energy spectrum [second term in the right-hand-side of Eq.\ (\ref{spec})] 
as a function of the magnetic flux through the 
ring, $\Phi/\Phi_0$, for the symmetry-restored (stationary) states $\Phi^{\rm PROJ}_L$ for $N=7$ fermionic ions. The 
remaining parameters are: Wigner parameter $R_W=1000$, ring radius $R=200 l_0$, and oscillator strength $l_0=50$ nm. 
According to Table \ref{t1} in the Appendix \ref{appa}, the parameter $C_R$ in Eq.\ (\ref{spec}) was taken equal to 
$1.7847 \times 10^{-6} \hbar \omega_0$. Each curve is labeled with the corresponding magic total angular momentum
$L_m$. The circles highlight several energy-crossing points most susceptible to symmetry breaking. 
Energies in units of $10^{-4} \hbar \omega_0$. 
}
\label{fspec}
\end{figure}

We have carried 
\textcolor{black}{out}
numerical calculations to determine the rotational spectrum of the $\Phi^{\rm PROJ}_L$'s.
For the calculation of $h(\gamma)$ and $n(\gamma)$, we use the rules for determinants composed of nonorthogonal 
orbitals; see, e.g., Ref.\ \cite{hurlbook}. Similar rules apply for permanents. The numerical calculations
are facilitated by the fact that the one-body and two-body matrix elements between the orbitals 
$u({\bf r}, {\bf R}_j)$ have closed analytic expressions \cite{ishi03,boys50,yannpub}. 

In all three cases [(i) odd number of repelling fermions, (ii) even number of repelling fermions, and (iii) 
attractive bosons], and for all values of $N \leq 10$, large localization parameters $R_W \geq 200$ and 
$R_\delta \geq 50$, and large ratios $R/l_0 \geq 40$ that we studied, we found that indeed the numerically calculated 
energies of the $\Phi^{\rm PROJ}_L$'s according to Eq.\ (\ref{eprj}) (see, e.g., Tables \ref{t1}, \ref{t2}, and 
\ref{t3} in Appendix \ref{appa}) can be well-fitted by that of an Aharonov-Bohm-type 
spectrum associated with a quantum many-body rigid rotor (see also \cite{li12,robi15}), i.e., 
\begin{align}
E^{\rm PROJ}(L) \approx V_{\rm int} + C_R (L - N \Phi/\Phi_0)^2.
\label{spec}
\end{align}   

\textcolor{black}{
$V_{\rm int}$ approximates the ground-state energy of the few-particle system and takes different values for 
different many-body wave functions.}

\textcolor{black}{The numerically determined coefficient $C_R$ is essentially a constant (see below).}
$\Phi=\pi R_{\rm eq}^2 B$ is the magnetic flux through the ring and $\Phi_0=h/\eta$ is the magnetic flux quantum. The 
values of the angular momenta $L$ are not arbitrary. Because of the crystalline symmetries, as well as the symmetric 
or antisymmetric behavior under particle exchange, they are given by proper sequences of magic angular momenta $L_m$ 
(see section \ref{symm} below for further discussion). In particular, all values of angular momenta are allowed for 
the case of attractive bosons, i.e., $L_m=0,\pm 1,\pm 2,\ldots$. For the case of fully polarized repelling fermions 
with $N$ odd or spinless repelling bosons with any $N$, the allowed angular momenta are restricted to the sequence 
$L_m=kN$, with $k=0,\pm 1,\pm 2,\ldots$. 
For the case of fully polarized repelling fermions with $N$ even, the allowed angular momenta are given by a different 
sequence $L_m=(k+1/2)N$, with $k=0,\pm 1,\pm 2,\ldots$. 

\textcolor{black}
{Due to the very large values of $R_W$ and $R_\delta$}, the value of $C_R$ is very close to that of a 
classical rigid rotor, corresponding to $N$ point particles in their equilibrium configuration inside the annular 
confinement, i.e., $C_R \approx C_R^{\rm cl}=\hbar^2/[2 {\cal I}(R_{\rm eq})]$, with inertia moment
${\cal I}(R_{\rm eq})=NM R_{\rm eq}^2$. As typical examples, in Table \ref{t1}, Table \ref{t2}, and Table \ref{t3} 
of Appendix \ref{appa}, we list calculated energies according to Eq.\ (\ref{eprj}) for an odd number $N=7$ and
an even number $N=8$ of fermionic ions, as well as for $N=10$ attractive bosons, respectively; $R_W=1000$ 
for the repelling ions and $R_\delta=50$ for the attractive bosons. The ratio $\tilde{f} \equiv C_R/C_R^{\rm cl} 
\approx 1$ for all $L < 165$ for ions, and for all $L \leq 30$ for attractive bosons. 
\textcolor{black}
{As aforementioned, the rigid-rotor-type spectrum in Eq.\ (\ref{spec}) was explored earlier in the 
QSTC literature \cite{li12,robi15}; however, by itself it does not lead to the derivation of appropriate QSTC wave 
functions. The demonstrated agreement between the microscopically calculated rotational part of the spectrum 
[Eq.\ (\ref{eprj})] and the analytic second term in Eq.\ (\ref{spec}) expected for a QSTC \cite{li12,robi15} validates
the expressions $\Phi^{\rm PROJ}_L$ introduced in Eq.\ (\ref{prj}) as proper trial wave functions for the QSTC.}

\textcolor{black}{
We showed previously that the limit of a quantum rigid rotor for a system of strongly interacting particles can 
also be reached in external confinements with geometries other than the ring geometry. In particular, the rigid-rotor
limit for $R_W=200$ and in a fully two-dimensional parabolic confinement was demonstrated for two electrons in Ref.\ 
\cite{yann00} using exact many-body wave functions and for a few electrons in Ref.\ \cite{yann04} using the
same ansatz as in Eq.\ (\ref{prj}) here. In this universal rigid-rotor limit, the rotational part of the spectra is
naturally similar. However, the presence of strong many-body correlations (which result from the 
beyond-mean-field, multi-determinant, or multi-permanent, nature of the wave function) is reflected in the actual 
numerical values of the first term, $V_{\rm int}$, in Eq.\ (\ref{spec}). 
}

\textcolor{black}{
In our scheme, which allows for an expanded variational freedom by employing unrestricted orbitals at the mean-field 
single-determinant, or single-permanent, level (i.e., a different orbital for each particle), the ground-state
energy is lowered at every step (see in particular Fig.\ 1 in Ref.\ \cite{yann07}): restricted HF (symmetry conserving)
$\rightarrow$ unrestricted HF (symmetry breaking) $\rightarrow$ symmetry restoration via projection techniques (an 
example of the full scheme can be seen in Fig.\ 5 of Ref.\ \cite{yann07}). This scheme for repelling bosons
translates as: symmetry-conserving Gross-Pitaevskii $\rightarrow$ symmetry-breaking ansatz of unrestricted 
permanent $\rightarrow$ symmetry restoration via projection techniques. We note that allowing the single orbital of the
Gross-Pitaevskii equation to break the azimuthal symmetry of the ring leads \cite{kana08,kana09,kana10} to 
higher-energy solitonic Bose-Einstein condensate branches in the rotational part of the spectrum, which are sharply 
different from the QSTC wave functions introduced in this paper.
}

\textcolor{black}{
As a specific example of the lowering of the ground-state energy in our scheme, we report that the energy for the 
unrestricted ansatz determinant [see Eq.\ (\ref{euhf})] in the case of $N=8$ ultracold ions on a ring of
radius $R=200l_0$ with $R_W=1000$ (case described in Table \ref{t2} of Appendix A) is $E_{\rm UHF}=118.1771$ 
$\hbar \omega_0$, while the corresponding restored-symmetry ground state has indeed a lower energy $V_{\rm int}=
117.9271$ $\hbar \omega_0$. This lowering of the total energy is immense compared to the quantum of the
rotational motion $C_R^{\rm cl}=1.5614 \times 10^{-6} \hbar \omega_0$ (see caption of TABLE \ref{t2} in Appendix A).} 

An illustrative case of the 
\textcolor{black}{rigid-rotor rotational spectra encoded in the second term} 
in Eq.\ (\ref{spec}) is displayed in Fig.\ \ref{fspec}.
A main feature of these spectra are the crossing points (several of them encircled) between pairs of curves
with different $L_m$'s. The crossings define special magnetic-field values, $\tilde{\Phi}/\Phi_0=(L_1+L_2)/(2N)$, 
in the neighborhood of which the system is particularly susceptible to symmetry breaking via the 
intermixing of two angular momenta and the ensuing generation of the PWM wave packets [see Eq.\ (\ref{pwm})].

Because the symmetry-restored (projected) wave function $\Phi^{\rm PROJ}$ [Eq.\ (\ref{prj})] preserves the 
group-theoretical requirements of the continuous 2D rotational group, its single-particle density is azimuthally 
uniform. However, the crystalline order of the original MF (symmetry-broken) wave function $\Psi^{\rm SB}$ is not 
destroyed in the symmetry-restoration step; instead, it mutates into a hidden order, which however
can be revealed via the conditional probability distribution (CPD) (density-density correlation function).
The CPD is  given by
\begin{align}
& {\cal D}({\bf r},{\bf r}_0) = \langle \Phi^{\rm PROJ}_L|\sum_{i\neq j} 
\delta({\bf r}_i,{\bf r}) \delta({\bf r}_j,{\bf r}_0)|\Phi^{\rm PROJ}_L \rangle.
\label{cpd}
\end{align}
The CPD provides the probabilty of finding a particle in position ${\bf r}$ assuming that another one
is located at the fixed point ${\bf r}_0$. Substitution of the expression [Eq.\ (\ref{prj})] that
defines $\Phi^{\rm PROJ}_L$, yields for ${\cal D}({\bf r},{\bf r}_0)$ a double integral over the azimuthal angles 
$\gamma_1$ and $\gamma_2$; this integral expression is given in the Appendix \ref{appb}. 

\begin{figure}[t]
\centering\includegraphics[width=8cm]{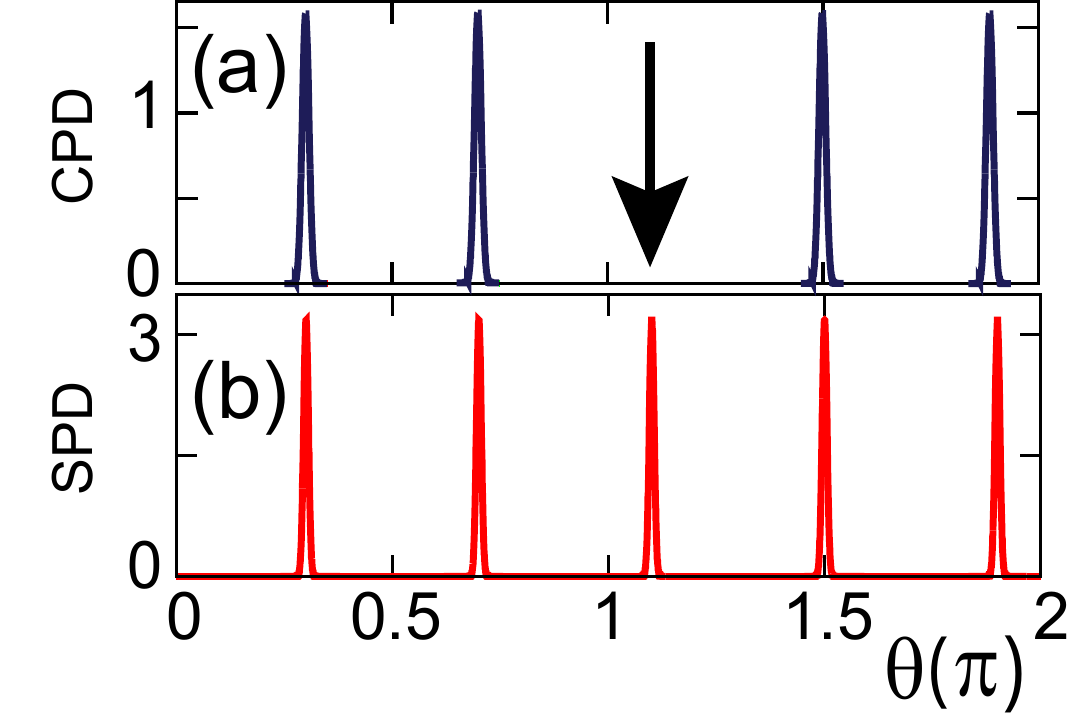}
\caption{
(a) CPD of the symmetry-restored stationary (beyond MF) state $\Phi^{\rm PROJ}_L$ for $N=5$ fermionic ions along the 
perimeter of the ring (at a radius $R_{\rm eq}$). The arrow at $\theta_0=1.1\pi$ denotes the fixed point 
${\bf r}_0=R_{\rm eq} e^{i \theta_0}=R_{\rm eq} e^{1.1 i \pi}$. Note the $2\pi/5$ angle between the nearest-neighbor 
humps and between the arrow and the two adjacent humps. Other parameters are: magic angular momentum
$L=L_m=N$, $R_W=200$, $R=40 l_0$, $l_0=50$ nm, 
and $\Phi/\Phi_0=0.8$. There is no hump at the fixed point. CPD in units of $10^{-2}/(2\pi \lambda^4)$.
(b) SPD of the original MF state $\Psi^{\rm SB}(\gamma=0.1\pi)$ (a determinant) for $N=5$ fermionic ions,
exhibiting explicitly symmetry breaking. Other parameters are $R=40 l_0$, $l_0=50$ nm.
SPD in units of $10^{-1}/(2\pi \lambda^2)$. In contrast to the symmetry-broken MF SPD in (b), the SPD of the 
symmetry-restored, beyond-mean-field $\Phi^{\rm PROJ}_L$ is azimuthally uniform; see black dashed line in Fig.\ 
\ref{fdens}(a). Azimuthal angle $\theta$ in units of $\pi$.
}
\label{fcpd}
\end{figure}

\begin{figure*}[t]
\centering\includegraphics[width=15cm]{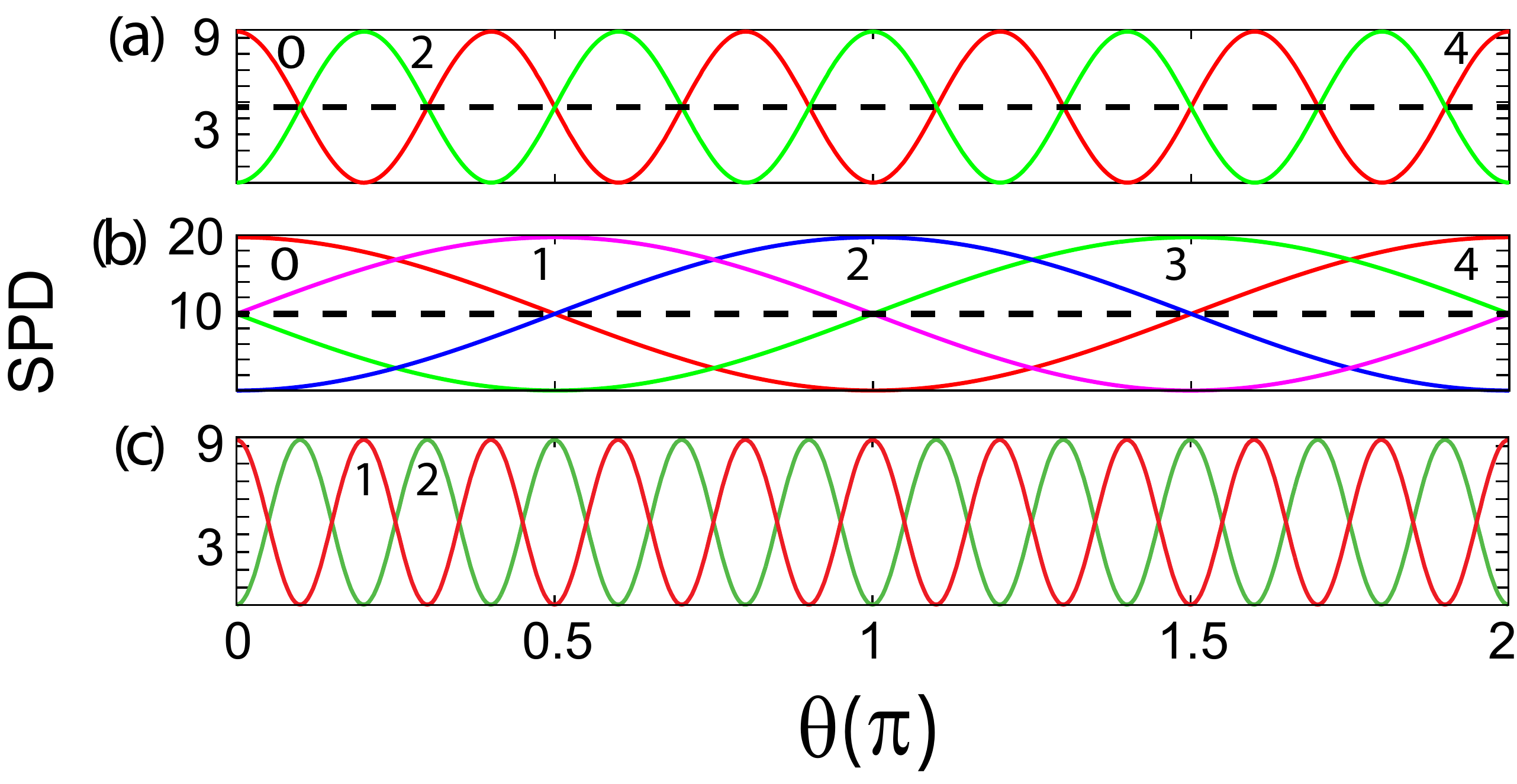}  
\caption{
Snapshots of undamped inhomogeneous single-particle densities rotating around the ring that were calculated with 
the wave packet $\Phi^{\rm PWM}(L_1,L_2; t)$. (a) $N=5$ fermionic ions and magic angular momenta $L_1=0$, $L_2=N$. 
The SPDs are shown for the time instances $t_j=j\tau/4$, with $j=0$, $j=2$, and $j=4$
(the $j$'s label the curves); $\tau= 2\pi \hbar /(|E^{\rm PROJ}(L_2)-E^{\rm PROJ}(L_1)|)$. 
Other parameters: $\alpha=\beta=1/\sqrt{2}$, $R_W=200$, 
$R=60 l_0$, $l_0=50$ nm, and $\Phi/\Phi_0=1.8$. 
(b) $N=7$ attractive bosons and magic angular momenta $L_1=0$, $L_2=1$. The SPDs are shown for the time instances 
$t_j=j\tau/4$, $j=0,1,\ldots,4$. Other parameters: $\alpha=\beta=1/\sqrt{2}$, $R_\delta=50$, $R=40 l_0$, $l_0=1$ 
$\mu$m, and $\Phi/\Phi_0=6.4$. The black horizontal dashed lines represent the uniform density of either one of the 
stationary states $\Phi^{\rm PROJ}_{L_m}$ (with $L_m=L_1$ or $L_m=L_2$) that contribute to the nonstationary wave 
packet $\Phi^{\rm PWM}(L_1,L_2; t)$. 
(c) $N=5$ fermionic ions and magic angular momenta $L_1=0$, $L_2=2N$ (higher-harmonic of the QSTC).  $t_j=j\tau/4$, 
with $j=1$ and $j=2$.
Other parameters as in (a). For repelling ions, panels (a) and (c), the period is $T=\tau/N$; for attractive bosons,
the period is $T=\tau$. The SPDs are in units of $10^{-2}/(2\pi\lambda^2)$. Azimuthal angle $\theta$
in units of $\pi$.}
\label{fdens}
\end{figure*}

Fig.\ \ref{fcpd}(a) displays an illustrative example of the hidden order in the symmetry-restored wave function
$\Phi^{\rm PROJ}_L$. The CPD in Fig.\ \ref{fcpd}(a) exhibits well localized features; it contrasts with the uniform 
horizontal black dashed lines in Figs.\ \ref{fdens}(a) and \ref{fdens}(b) which describe $\rho({\bf r},t)$'s of 
$\Phi^{\rm PROJ}_L$ along the perimeter of the ring trap (at a radius $R_{\rm eq}$).
\textcolor{black}{
We stress that the fixed point ${\bf r}_0$ in the CPD is arbitrary, i.e., the four peaks in the CPD in Fig.\ 
\ref{fcpd}(a) readjust to a different choice of $\theta_0$ so that the relative distance between them and the arrow 
remain unchanged.}
Fig.\ \ref{fcpd}(b) displays the SPD of the original state $\Psi^{\rm SB}(\gamma=0.1\pi)$ 
(a determinant) for $N=5$ fermionic ions, exhibiting explicitly the symmetry breaking at the mean-field level. 

\subsection{Third level (beyond mean field): Periodic time evolution of the spatially inhomogeneous 
$\rho({\bf r},t)$ associated with the wavepacket $\Phi^{\rm PWM}(L_1,L_2; t=0)$}
\label{tl}

As aforementioned, the two-state wave packet in Eq.\ (\ref{pwm}) is not an eigenstate of the total angular 
momentum, and thus it is not a stationary state when the pinning agent is lifted; such a pinning agent could be 
implemented, for example, as a distortion of the circular geometry of the trap confinement, or as a modulation of 
the trap potential in the azimuthal direction along the ring \cite{haef17}. (A sudden variation of the magnetic 
field can also transform an eigenstate $\Phi^{\rm PROJ}_L(B_1)$ at a given $B_1$ value to a superposition of 
$\Phi^{\rm PROJ}_L(B_2)$ states at another $B_2$ value \cite{yannpub}.) The resulting time 
evolution is associated with a time-dependent phase $\phi(t)$ as discussed previously. Here we will show 
explicitly that $\phi(t)$ represents an undamped rotation of spatially inhomogeneous $\rho({\bf r},t)$'s
around the ring, so that the many-body $\Phi^{\rm PWM}(L_1,L_2; t)$ exhibit the desired behavior of a QSTC. The 
successful theoretical identification and experimentally implemented superposition of two appropriate many-body 
spin eigenstates of the Ising Hamiltonian (resulting in a ``spin Schr\"{o}dinger-Cat'' state) were  
keys to the emulation of the ``weaker class'' of discrete time crystals \cite{sond16,naya16,monr17,fazi17}.

The $\rho({\bf r},t)$ of $\Phi^{\rm PWM}(L_1,L_2; t)$ is defined as
\begin{align}
\rho({\bf r};t) = \langle \Phi^{\rm PWM}(L_1,L_2; t) | \sum_{i=1}^N \delta({\bf r}_i-{\bf r}) | 
\Phi^{\rm PWM}(L_1,L_2; t) \rangle.
\label{rho}
\end{align} 

\textcolor{black}{As in the case of the CPD,}
$\rho({\bf r};t)$ entails a double integral over the azimuthal angles $\gamma_1$ and $\gamma_2$; 
the lengthy expression is given in Appendix \ref{appc}.

Fig.\ \ref{fdens} displays the periodic time evolution of $\rho({\bf r},t)$'s for two illustrative 
$\Phi^{\rm PWM}(L_1,L_2; t)$ cases, one for $N=5$ Coulomb repelling fermionic ions [Fig.\ \ref{fdens}(a)] with 
$L_2-L_1=N$ and the other for $N=7$ neutral bosons with $L_2-L_1=1$ [Fig.\ \ref{fdens}(b)] interacting via an 
attractive contact interaction. The $\rho({\bf r},t)$'s were calculated at times $t_j=j\tau/4$, where 
$\tau = 2 \pi \hbar /(|E^{\rm PROJ}(L_2)-E^{\rm PROJ}(L_1)|)$; 
\textcolor{black}{the actual used} 
$j$'s label the
$\rho({\bf r},t)$ curves. The number of humps exhibited by the PWM $\rho({\bf r},t)$'s in Fig.\ \ref{fdens}(a) 
and Fig.\ \ref{fdens}(b) is equal to that in the original MF densities, i.e., $N$ for the repelling-fermions PWM 
and one for the attractive-bosons lump. The period of the PWM $\rho({\bf r},t)$'s is $T=\tau/N$ for repelling ions and
$T=\tau$ for attractive bosons. 

Finally, Fig.\ \ref{fdens}(c) demonstrates a different state of matter, i.e., multi-harmonic excitations of the QSTC
exhibiting a multiple number of density humps, i.e., $k N$ and $k$ (with $k=2,3,\ldots$), 
corresponding to $\Phi^{\rm PWM}(L_1,L_2; t)$'s with $L_2-L_1=k N$ for repelling
fermions and with $L_2-L_1 =k$ for attractive bosons, respectively.

\textcolor{black}{
We note that the PWM broken-symmetry state introduced here to describe a QSTC has an energy intermediate between 
$E_1$ and $E_2$ because $\alpha^2+\beta^2=1$ (i.e., $E^{\rm PWM}=\alpha^2 E_1 + \beta^2 E_2$). In particular, at the 
crossing point of the two parabolas (where $E_1=E_2=E_{\rm cross}$), one has always $E^{\rm PWM}=E_{\rm cross}$. This
contrasts with the behavior of the energy of the non-crystalline states studied in Refs.\ \cite{kana08,kana09,kana10},
which lies always well above the crossing point.}

\section{Discussion}
\label{disc}

\subsection{Symmetries of the trial wave functions, magic angular momenta, and rigidity}
\label{symm}

\textcolor{black}{
Despite the fact that the trial wave functions in Eq.\ (\ref{prj}) are a 
good approximation to the rotational-symmetry-preserving 
many-body eigenstates, they do embody and reflect in an optimum way the crystalline point-group symmetries (familiar 
from bulk crystals).} 
Specifically, the $C_N$ point-group symmetry of the ``classical'' crystal, which is accounted for
through the kernel of symmetry-broken MF determinants (or permanents) $\Psi^{\rm SB}$, is reflected in the fact that
the trial wave functions $\Phi^{\rm PROJ}_L$ are identically zero except for a subset of {\it magic\/} angular 
momenta $L_m$. 

In the case of $N$ repelling particles, the magic total angular momenta can be determined by considering the 
point-group symmetry operator $\hat{R}(2\pi/N) \equiv \exp (-i 2\pi {\hat L} /N)$ that rotates on the ring 
simultaneously the localized particles by an angle $2\pi/N$. 
\textcolor{black}{
In connection to the state $\Phi^{\rm PROJ}_L$, the operator $\hat{R}(2\pi/N)$ can be invoked in two different ways, 
namely either by applying it on the ``intrinsic'' part $\Psi^{\rm SB}$ or the ``external'' phase factor 
$\exp(i \gamma L)$ (see Ch. 4-2c Ref.\ \cite{bomo98}).}
One gets in the case of fermions
\begin{equation}  
\hat{R}(2\pi/N) \Phi^{\rm PROJ}_L = (-1)^{N-1} \Phi^{\rm PROJ}_L,
\label{rot1} 
\end{equation}
from the first alternative and
\begin{equation}  
\hat{R}(2\pi/N) \Phi^{\rm PROJ}_L = \exp(-2\pi L i/N) \Phi^{\rm PROJ}_L,
\label{rot2} 
\end{equation}
from the second alternative. The $(-1)^{N-1}$ factor in Eq.\ (\ref{rot1}) results from the fact that the $2\pi/N$ 
rotation is equivalent to exchanging $N-1$ rows in the  $\Psi^{\rm SB}$ determinant. Now if 
$\Phi^{\rm PROJ}_L \neq 0$, the only way that Eqs.\ (\ref{rot1}) and (\ref{rot2}) can be simultaneously true is
if the condition $\exp (2\pi L i/N)=(-1)^{N-1}$ is fulfilled. This leads to the following sequence of magic angular 
momenta,
\begin{equation}
L_m = k N; \;\;\; k=0,\pm 1, \pm 2, \pm 3, \ldots,
\label{mag1}
\end{equation}
for $N$ odd, and
\begin{equation}
L_m = (k + \frac{1}{2}) N; \;\;\; k=0, \pm 1,\pm 2,\pm 3, \ldots,
\label{mag2}
\end{equation}
for $N$ even.

Because a permanent is symmetric under the interchange of two rows, the corresponding magic $L_m$'s for 
spinless bosons are given by the sequence in Eq.\ (\ref{mag1}) for both odd and even numbers of localized bosons. 

Regarding the numerical aspects, the fact that $\Phi^{\rm PROJ}_L$ is zero for non-magic $L$ values results in the 
\textcolor{black}{vanishing} 
(within machine precision) of the normalization factor $\int_0^{2\pi} n(\gamma)e^{i\gamma L} d\gamma$ 
in Eq.\ (\ref{eprj}). As a result only the physically meaningful energies associated with 
magic angular momenta are given in Table \ref{t1}, Table \ref{t2}, and Table \ref{t3} of Appendix \ref{appa}.

\textcolor{black}{
We stress that the properties and physics associated with magic-angular-momentum yrast states are well known in the 
literature of 2D quantum dots \cite{yann07,ruan95,maks96,seki96,maks00}. 
Of immediate relevance to this paper is the enhanced energy stabilization that they acquire
in their neighborhood (thus characterized often as ``cusp'' states) in the regime of strong interactions (i.e., for
large $R_W$ or $R_\delta$). This energy stabilization can be explicitly seen in Fig.\ 15 of Ref.\ \cite{yann07}, where
the triplet state correponds to the fully polarized case for two electrons with magic angular momenta
$L_m=(2k+1)$, $k=0,\pm 1, \pm 2, \ldots$. The fact that large energy gaps do develop between the 
magic-angular-momentum rotational yrast states and the other (excited) states is also well established in the QSTC
literature; for the case of ultracold ions on a ring, see Refs.\ \cite{li12,robi15}, and for the case
of the bosonic lump, see Ref.\ \cite{wilc12}.}

In this paper, we consider fully polarized fermions only, that is cases when $S=S_z=N/2$, where $S$ is the
total spin and $S_z$ is its projection. Consideration with our methodology of the other spin values $S_z < N/2$ 
is straightforward; it requires, however, restoration of both the total spin ${\bf S}^2$
and the total angular momentum. An explicit example for $N=3$ fermions is discussed in Ref.\ \cite{yann03}.

In addition to the magic angular momenta,
the properties of the original crystalline structure
\textcolor{black}{built-in}
in $\Psi^{\rm SB}$ are reflected in the high-degree of
rigidity exhibited by the symmetry-preserving $\Phi^{\rm PROJ}_L$. As demonstrated previously, the SPD of
$\Phi^{\rm PROJ}_L$ is uniform, but the CPD of 
$\Phi^{\rm PROJ}_L$ reveals the now hidden crystalline structure of $N$ strongly repelling  particles on the ring. 
The rigidity of $\Phi^{\rm PROJ}_L$ is manifested in that the CPDs have the same $N$-hump shape and are 
independent of the actual value of the magic angular momenta, as well as of the fermion or boson statistics and
of whether the number $N$ of fermions is odd or even. 
This rigidity is a consequence of the strong two-body interaction and cannot be found in many-body
wave functions associated with weak interparticle interactions. 

\subsection{Initial wave packets and associated time evolution}

The focus of this paper is the construction of a symmetry-preserving wave function $\Phi^{\rm PROJ}_L$ associated with
the finite-crystal symmetry-broken determinant (or permanent) $\Psi^{\rm SB}$. However, it is instructive to inquire
about the reverse process, that is how to represent the symmetry-broken crystal as a superposition in the
complete basis formed by the symmetry-preserving $\Phi^{\rm PROJ}_L$'s. Indeed one can write the expansion
\begin{equation}
\Psi^{\rm SB} = \sum_L {\cal C}_L \Phi^{\rm PROJ}_L,
\label{sbe}
\end{equation} 
where [using Eq.\ (\ref{prj})] the expansion coefficients are given by
\begin{equation}
{\cal C}_L = \frac{1}{2\pi} \int_0^{2\pi} d\gamma e^{-i\gamma L} n(\gamma),
\label{sbc}
\end{equation} 
and the norm overlap $n(\gamma)$ was defined in Eq.\ (\ref{nov}). Of course the index $L$ runs over the
appropriate sequence of magic angular momenta as discussed in Section \ref{symm}.

Eqs.\ (\ref{sbe}) and (\ref{sbc}) illustrate the fact that with respect to the exact many-body (linear)
Schr\"{o}dinger equation the symmetry-broken-crystal wave 
function $\Psi^{\rm SB}$ is a wave packet and not a stationary eigenstate. This is also in general true for all 
SB mean-field solutions, whether they are solutions of the unresticted Hartree-Fock equations in the case of confined
electrons (e.g., in quantum dots \cite{yann07}), or they are the familiar solitonic solutions of the 
Gross-Pitaevskii equations in onedimensional bosonic systems. Due to the rapid experimental control, 
the latter cases are currently attracting a 
lot of attention. Indeed motivated by experiments that suggest the need to go beyond-mean field dynamics, the number
of related theoretical investigations has burgeoned \cite{degu12,kavo17,cede11,carr16,schm17}.

These theoretical studies investigate how an initial state approximating a solution of the nonlinear 
Gross-Pitaevskii equation evolves in time under the
exact many-body Hamiltonian. For both the cases of dark \cite{degu12,kavo17} (a hole in matter density) 
and bright \cite{cede11,carr16,schm17} solitons (an excess in matter density, like the case of the lump considered 
in this paper), these studies are finding a ``universal'' behavior of dispersion in space, decay in time, and 
time revival 
\textcolor{black}{at the initial position} 
of the soliton. This behavior can be understood by taking into consideration the expansion in Eq.\ (\ref{sbe}).
In fact, each energy mode in 
\textcolor{black}{Eq.\ (\ref{sbe})} 
will evolve in time according to its own phase $\exp(-iE_L t/\hbar)$, 
and the interaction between all of them results in a decay-type behavior. Moreover, the initial occupation
amplitudes ${\cal C}_L^2$ of the different modes are unequal, 
\textcolor{black}{and as a result} 
probability flows from the
higher occupied modes to those with lower initial occupations, which leads to a dispersive behavior.
However, because the system is finite, there exists a Poincar\'{e} period, and the system will eventually
experience a revival \cite{bocc57,schu78}.

For achieving a QSTC, we propose here a different initial wave packet, i.e., a two-mode one with equal weights, as 
specified in Eq.\ (\ref{pwm}). As explicitly demonstrated through numerical claculations, such a two-mode initial wave 
packet preserves at all times, and without damping, the spatial and temporal periodicities
expected from the classical finite crystal. We note that the consideration of two-mode Schr\"{o}dinger-cat 
states is a key element in the theory of the discrete time crystal, where the focus is to enable a sloshing
behavior between these specialized paired states by minimizing interactions to the rest of the system. In fact,
our symmetry-preserving trial functions $\Phi^{\rm PROJ}_L$ [see Eq.\ (\ref{prj})] can be viewed as a more
complex class of Schr\"{o}dinger-cat states. This analogy is straightforward for the case of the mirror superposition
used in Ref.\ \cite{sach15}, which can be reproduced from expression (\ref{prj}) as a limiting case by using
only two angles $\gamma_1=0$ and $\gamma_2=\pi$.

\subsection{Relation to configuration-interaction (CI) wave functions}

As mentioned previously, the symmetry-preserving trial functions $\Phi^{\rm PROJ}_L$ are identically zero
for values of $L$ different from the magic angular momenta $L_m$; see section \ref{symm}. Naturally the
exact many-body spectrum has a plethora of additional states with good $L$, which however cannot be reached with the
approach in this paper. Indeed this approach is tuned to extracting from the complete spectrum only the ground
states that correspond to non-vibrating classical finite crystal arrangements. The remaining many-body states can be
reached by using the CI approach, which is in principle an exact methodology when converged; the CI is often 
referred to as exact diagonalization (EXD). The CI approach is computationally expensive, but comparisons between 
the symmetry-restored trial functions and the CI wave functions have been used by us to demonstrate the numerical 
accuracy of the symmetry-restored wave functions, as well as to clarify their special place in the whole spectrum, 
namely that for particular magnetic-field ranges they can become the global ground state, as is the case with
the Aharonov-Bohm spectrum in Fig.\ \ref{fspec} of the present paper. Higher-in-energy CI 
solutions with different $L$ (and also with $L=L_m$) do incorporate vibrational and other type of internal 
excitations, and as a result a superposition of two random CI states with good $L_1$ and $L_2$ will not necessarily 
exhibit the crystalline single-particle-density structure of exactly $N$ humps.

Systematic comparisons between symmetry-restored states and CI wave functions have been carried out by us
previously for the case of a few electrons confined in parabolic quantum dots. Although the external confining
potential and particle species in parabolic QDs are different from the case of the ring traps considered here, the 
symmetry properties of the many-body wave functions are universal. Thus the analysis presented in our previous QD 
studies can be used to gain further insights to the results for the QSTC presented in section \ref{sbsp}. 
In particular, Fig.\ 6 and Fig.\ 7 of Ref.\ \cite{yann11} offer an explicit illlustration of the fact 
that $N$-humped crystalline SPD structures arise only when both $L_1$ and $L_2$ coincide with magic angular
momenta and the associated CI wave functions correspond to global ground states in some range of magnetic 
fields (see Fig.\ 5 and section III in Ref.\ \cite{yann11}).

In our previous studies of QDs, excellent agreement was found between the total energies of symmetry-restored trial 
functions and the corresponding CI energies for both the cases with or without an applied magnetic field, as 
testified by the many reported direct numerical comparisons. We mention here a few specific examples, i.e., 
Table III and IV in Ref.\ \cite{yann03}, Table IV in Ref.\ \cite{yann03.2}, and Fig.\ 4 in Ref.\ \cite{yann09}.  

Such systematic numerical comparisons between symmetry-restored and CI wave functions for the case of 
ring-trapped ultracold ions and neutral bosons are outside the scope of the present paper. However, they will
be reported in subsequent publications \cite{yannpub}, including the case away from the quasi-1D regime 
(when the dependence on the ring width becomes important). 

\section{Conclusions}
\label{concl}

The discussion \cite{wilc13,li13,wata15}, motivated by the criticism \cite{brun13,brun13.2,nozi13,wata15} of the 
original \cite{wilc12,li12} QSTC proposals (which were based on ground states), spurred speculations about 
non-equilibrium low-lying states as possible instruments for describing QSTCs. 
For $N$ rotating particles on a ring, and using the theory of 
symmetry breaking and symmetry restoration via projection techniques \cite{yann07}, this
paper succeeded in explicitly uncovering the existence of low-lying states with QSTC behavior, by introducing 
beyond-MF appropriate trial many-body wave functions (see Fig.\ \ref{fdens}). Along with its conceptual and 
methodological significance, we propose to focus experimental attention on selected applied magnetic field values 
where the Aharonov-Bohm-type spectra corresponding to different magic angular momenta are most susceptible to mixing 
(Fig.\ \ref{fspec}), resulting in rotating pinned-Wigner-molecule many-body states found here to exhibit QSTC 
behavior. This constructive platform fills an apparent gap in the quest for ultracold ring-confined ions or 
neutral-atom QSTCs. 

We recall that although the original proposals for the quantum space-time crystal \cite{wilc12,li12} suggested 
realization of the concept through the use of ultracold few ring-trapped particles, this is yet to be achieved 
experimentally. Nevertheless, for the case of ultracold ions, several publications have reported significant progress 
in controlling aspects of a quantum rotor on a ring. In particular, the ability 
to generate and control symmetry-breaking through pinning of the rotating ion crystal has been demonstrated 
by using up to 15 $^{40}$Ca$^+$ ions in a ring with a microfabricated 
silicon surface Paul trap \cite{haef17}, or 3 $^{40}$Ca$^+$ ions in a 2D ring-type configuration in a linear
Paul trap \cite{nogu14}. To fully inplement and control the QSTC trial functions presented in this paper, the ions 
need to be cooled 
\textcolor{black}{down to near} 
the ground states. In this respect, Ref.\ \cite{haef17} has achieved temperatures 
$\sim$ 3 mK (for a trap with a radius of $\sim$ 60 $\mu$m), while Ref.\ \cite{nogu14} reported temperatures
in the nanometer range (for an effective ring radius in the 6 to 8 
\textcolor{black}{micrometer range).} 
It is expected that cooling techniques
and procedures will be further optimized and will be successful in the near future in producing near-ground-state 
temperatures, as is exemplified by a very recent publication \cite{che17}. 

An essential requirement, met by our theory, is that it is imperative that the proposed beyond-mean-field many-body  
trial wave functions (i.e., beyond the UHF or GP treatments) for predicting proper quantum space-time-crystal 
behavior of particles moving on a ring will be based on solutions to the 
interacting particles Schr\"{o}dinger equation that posses good angular momenta, as well as exhibit (hidden) 
ordering that reflects an underlying finite crystalline symmetry. This is achieved in our theory 
through the first two construction stages, namely, the unrestricted Hartree-Fock solution followed by an 
angular momentum projection, yielding the function $\Phi^{\rm PROJ}_L$ [Eq.\ (\ref{prj})]. It is then proposed by us 
that these projected and stationary many-body wave functions are susceptible to mixing, 
\textcolor{black}{see Eq.\ (\ref{pwm}),}   
favored to occur in the vicinity of crossings of Aharonov-Bohm-type spectra of ground-state energies {\it vs.\/} 
applied magnetic field (see circles in Fig.\ \ref{fspec}). This mixing results in non-stationary low-lying states that,
when evolved with the many-body Hamiltonian, yield undamped and non-dispersing periodic oscillations in both space and 
time.

\begin{acknowledgments}
Work supported by the Air Force Office of Scientic Research under Award No. FA9550-15-1-0519. 
Calculations were carried out at the GATECH Center for Computational Materials Science.
\end{acknowledgments}

\appendix
~~~~~~~~~\\
~~~~~~~~~\\
~~~~~~~~~\\
~~~~~~~~~\\
~~~~~~~~~\\
~~~~~~~~~\\
\begin{widetext}
\section{Numerical Calculations of the many-body rotational energies $E^{\rm PROJ}(L_m)$ [Eq.\ (\ref{eprj})]}
\label{appa}

Tables \ref{t1}, \ref{t2}, and \ref{t3} below present three illustrative examples of the rotational energy 
spectra $E^{\rm PROJ}(L_m)$ according to numerical calculations of the many-body expression in 
Eq.\ (\ref{eprj}) of the main text. The captions explain 
how the numerical $C_R$ in Eq.\ (\ref{spec}) is extracted from the computed values of $E^{\rm PROJ}(L_m)$.
$C_R$ is found to be very close to the classical rigid-rotor value 
$C_R^{\rm cl}=\hbar^2/[2 {\cal I}(R_{\rm eq})]$.

\begin{table*}[ht]
\caption{Rotational energy spectra according to Eq.\ (\ref{eprj}) and ratio $\tilde{f} \equiv C_R/C_R^{\rm cl}$ 
for $N=7$ spin polarized ultracold fermionic ions at two different magnetic fields $\Phi=0$ and $\Phi/\Phi_0=3.2$. The 
interparticle interaction is a repelling Coulomb potential. The energies are in units of 
$\hbar \omega_0=\hbar^2/(M l_0^2)$. The remaining parameters are: Wigner parameter $R_W=1000$, ring radius 
$R=200 l_0$, and oscillator strength $l_0=50$ nm. As a function of $L_m$, the numerically extracted coefficient 
$C_R$ in Eq.\ (\ref{spec}) was determined from the ratio 
$C_R=\left( E^{\rm PROJ}(L_m)-E^{\rm PROJ}(L_m-N) \right)/\left( N(2L_m-2 \Phi N/\Phi_0 - N) \right) $. 
Its value is practically constant and equal to $C_R^{\rm cl}$; see the values of the ratio $\tilde{f}$, which
are very close to unity. The classical rigid-body value is $C_R^{\rm cl}=1.7847 \times 10^{-6} \hbar \omega_0$. 
The underlined numbers refer to the ground state for a given $\Phi/\Phi_0$.} 
\begin{ruledtabular}
\begin{tabular}{cccccc}
\multicolumn{3}{c}{$N=7$ fermions, ~~~~$\Phi=0$ } & 
\multicolumn{3}{c}{$N=7$ fermions, ~~~~$\Phi/\Phi_0=3.2$} \\ \hline
$L_m$ & $E^{\rm PROJ}(L_m)$ & $\tilde{f}$ & $L_m$ & $E^{\rm PROJ}(L_m)$ & $\tilde{f}$ \\ \hline
\underline{0}  &  \underline{85.6564962006}  &  ~~~ &       0     & 85.6573927150 & ~~~ \\
7           &  85.6565836512  &  1.000011 &  7     & 85.6569201655 & 1.000681 \\
14          &  85.6568460029  &  1.000011 &  14    & 85.6566225173 & 1.001074 \\
21          &  85.6572832557  &  1.000011 &  \underline{21}    & \underline{85.6564997702} & 1.002594 \\
28          &  85.6578954097  &  1.000011 &  28    & 85.6565519241 & 0.993980 \\
35          &  85.6586824648  &  1.000011 &  35    & 85.6567789793 & 0.998619 \\
 $\ldots$   &  $\ldots$       &  $\ldots$ &  $\ldots$ & $\ldots$ & $\ldots$ \\
105         &  85.6761725766  &  1.000011 &  105   & 85.6686690913 & 0.999850 \\
112         &  85.6788835437  &  1.000010 &  112   & 85.6708200583 & 0.999863 \\
119         &  85.6817694118  &  1.000010 &  119   & 85.6731459265 & 0.999874 \\
126         &  85.6848301808  &  1.000010 &  126   & 85.6756466955 & 0.999884 \\
133         &  85.6880658508  &  1.000010 &  133   & 85.6783223655 & 0.999892 \\
140         &  85.6914764217  &  1.000010 &  140   & 85.6811729366 & 0.999899 \\
\end{tabular}
\end{ruledtabular}
\label{t1}
\end{table*}

\begin{table*}[ht]
\caption{Rotational energy spectra according to Eq.\ (\ref{eprj})
and ratio $\tilde{f} \equiv C_R/C_R^{\rm cl}$ for $N=8$ 
spin polarized ultracold fermionic ions at two different magnetic fields $\Phi=0$ and $\Phi/\Phi_0=3.2$. The 
interparticle interaction is a repelling Coulomb potential. The energies are in units of 
$\hbar \omega_0=\hbar^2/(M l_0^2)$. The remaining parameters are: Wigner parameter $R_W=1000$, ring radius 
$R=200 l_0$, and oscillator strength $l_0=50$ nm. As a function of $L_m$, the numerically extracted coefficient 
$C_R$ in Eq.\ (\ref{spec}) was determined from the ratio 
$C_R=\left( E^{\rm PROJ}(L_m)-E^{\rm PROJ}(L_m-N) \right)/\left( N(2L_m-2 \Phi N/\Phi_0 - N) \right) $. 
Its value is practically constant and equal to $C_R^{\rm cl}$; see the values of the ratio $\tilde{f}$, which
are very close to unity. The classical rigid-body value is $C_R^{\rm cl}=1.5614 \times 10^{-6} \hbar \omega_0$. 
The underlined numbers refer to the ground state for a given $\Phi/\Phi_0$.} 
\begin{ruledtabular}
\begin{tabular}{cccccc}
\multicolumn{3}{c}{$N=8$ fermions, ~~~~$\Phi=0$ } & 
\multicolumn{3}{c}{$N=8$ fermions, ~~~~$\Phi/\Phi_0=3.2$} \\ \hline
$L_m$ & $E^{\rm PROJ}(L_m)$ & $\tilde{f}$ & $L_m$ & $E^{\rm PROJ}(L_m)$ & $\tilde{f}$ \\ \hline
\underline{4}  &  \underline{117.9270981536}  &  ~~~ &       4     & 117.9278028694 & ~~~ \\
12          &  117.9272980155 &  1.000010 &  12    & 117.9273627314 & 1.001015 \\
20          &  117.9276977394 &  1.000010 &  20    & 117.9271224553 & 1.001852 \\
28          &  117.9282973253 &  1.000010 &  \underline{28} & \underline{117.9270820410} & 1.011065 \\
36          &  117.9290967731 &  1.000010 &  36    & 117.9272414889 & 0.997248 \\
44          &  117.9300960828 &  1.000010 &  44    & 117.9276007987 & 0.998782 \\
 $\ldots$   &  $\ldots$       &  $\ldots$ &  $\ldots$ & $\ldots$ & $\ldots$ \\
124         &  117.9510815852 &  1.000010 &  124   & 117.9421863015 & 0.999823 \\
132         &  117.9542793756 &  1.000010 &  132   & 117.9447440920 & 0.999837 \\
140         &  117.9576770279 &  1.000010 &  140   & 117.9475017443 & 0.999850 \\
148         &  117.9612745420 &  1.000010 &  148   & 117.9504592585 & 0.999860 \\
156         &  117.9650719179 &  1.000010 &  156   & 117.9536166343 & 0.999870 \\
164         &  117.9690691554 &  1.000010 &  164   & 117.9569738719 & 0.999878 \\
\end{tabular}
\end{ruledtabular}
\label{t2}
\end{table*}
~~~~~~~~~~~~~~~~\\
~~~~~~~~~~~~~~~~\\
~~~~~~~~~~~~~~~~\\
~~~~~~~~~~~~~~~~\\
~~~~~~~~~~~~~~~~\\

\begin{table*}[t]
\caption{Rotational energy spectra according to Eq.\ (\ref{eprj}) and ratio 
$\tilde{f} \equiv C_R/C_R^{\rm cl}$ for $N=10$ spinless 
ultracold bosons at two different magnetic fields $\Phi=0$ and $\Phi/\Phi_0=2.464$. The interpaticle contact
interaction is attractive. The energies are in units of $\hbar \omega_0=\hbar^2/(M l_0^2)$. The remaining 
parameters are: Wigner parameter $R_\delta=50$, ring radius $R=40 l_0$, and oscillator strength $l_0=1$ $\mu$m. As a 
function of $L_m$, the numerically extracted coefficient $C_R$ in Eq.\ (\ref{spec}) was determined 
from the ratio $C_R=\left( E^{\rm PROJ}(L_m)-E^{\rm PROJ}(L_m-1) \right)/ (2L_m-2 \Phi N/\Phi_0 - 1) $. Its value is 
practically constant and equal to $C_R^{\rm cl}$; see the values of the ratio $\tilde{f}$, which are very close to 
unity. The classical rigid-body value is $C_R^{\rm cl}=3.1250 \times 10^{-5} \hbar \omega_0$. The underlined 
numbers refer to the ground state for a given $\Phi/\Phi_0$.}
\begin{ruledtabular}
\begin{tabular}{cccccc}
\multicolumn{3}{c}{$N=10$ bosons, ~~~~$\Phi=0$ } & 
\multicolumn{3}{c}{$N=10$ bosons, ~~~~$\Phi/\Phi_0=2.464$} \\ \hline
$L_m$ & $E^{\rm PROJ}(L_m)$ & $\tilde{f}$ & $L_m$ & $E^{\rm PROJ}(L_m)$ & $\tilde{f}$ \\ \hline
\underline{0}  & \underline{-350.8488583900}   &  ~~~      &   0     & -350.8303108498  & ~~~ \\
1  & -350.8488271326   &  1.000234 &   1     & -350.8318195924  & 0.999995 \\
2  & -350.8487333607   &  1.000234 &   2     & -350.8332658201  & 0.999985 \\
3  & -350.8485770740   &  1.000234 &   3     & -350.8346495329  & 0.999973 \\
4  & -350.8483582728   &  1.000234 &   4     & -350.8359707308  & 0.999961 \\
5  & -350.8480769568   &  1.000234 &   5     & -350.8372294139  & 0.999947 \\
 $\ldots$   &  $\ldots$       &  $\ldots$ &  $\ldots$ & $\ldots$ & $\ldots$ \\
23 & -350.8323232897   &  1.000231 &   23    & -350.8491956906  & 0.997529 \\
24 & -350.8308542000   &  1.000231 &   24    & -350.8492665957  & 0.995160 \\
25 & -350.8293225964   &  1.000231 & \underline{25}    & \underline{-350.8492749867}  & 0.958962 \\
26 & -350.8277284788   &  1.000231 &   26    & -350.8492208634  & 1.006945 \\
27 & -350.8260718473   &  1.000230 &   27    & -350.8491042260  & 1.003333 \\
28 & -350.8243527019   &  1.000230 &   28    & -350.8489250745  & 1.002246 \\
29 & -350.8225710428   &  1.000230 &   29    & -350.8486834090  & 1.001722 \\
30 & -350.8207268699   &  1.000229 &   30    & -350.8483792296  & 1.001414 \\
\end{tabular}
\end{ruledtabular}
\label{t3}
\end{table*}
~~~~~~\\
~~~~~~\\

~~~~~~\\
~~~~~~\\

~~~~~~\\
~~~~~~\\

~~~~~~\\
~~~~~~\\

\section{Conditional Probability Distribution}
\label{appb}

The explicit expression for the CPDs of the symmetry-restored wave functions $\Phi^{\rm PROJ}_L$ 
[see Eq.\ (\ref{cpd})] is given by
\begin{align}
 {\cal D}({\bf r},{\bf r}_0) =
\frac{ \int_0^{2\pi} d\gamma_1 \int_0^{2\pi} d\gamma_2 e^{i(\gamma_1-\gamma_2)L}
\sum_{k\neq m,l\neq n} 
\left( {\cal G}^{ln}_{km}(\gamma_1,\gamma_2) \mp {\cal G}^{nl}_{km}(\gamma_1,\gamma_2) \right)
{\cal S}^{km}_{ln}(\gamma_1,\gamma_2)} {2 \pi \int_0^{2\pi} n(\gamma) e^{i\gamma L} d\gamma},
\label{cpd2}
\tag{A.1}
\end{align}
where 
\begin{align}
{\cal G}^{ln}_{km}&(\gamma_1,\gamma_2)= \frac{1}{\pi^2 \lambda^4} \exp \bigg( -\frac
{({\bf r}-{\bf R}_k(\gamma_1))^2 + ({\bf r}-{\bf R}_l(\gamma_2))^2 +
({\bf r}_0-{\bf R}_m(\gamma_1))^2 + ({\bf r}_0-{\bf R}_n(\gamma_2))^2}
{2 \lambda^2} \bigg) \times \nonumber \\
&\exp \bigg( i \frac
{ x(Y_k(\gamma_1)-Y_l(\gamma_2))+ y(X_l(\gamma_2)-X_k(\gamma_1))+
 x_0 (Y_m(\gamma_1)-Y_n(\gamma_2))+y_0(X_n(\gamma_2)-X_m(\gamma_1))}
{2 l_B^2} \bigg),
\label{gkern}
\tag{A.2}
\end{align} 
and the ${\cal S}^{km}_{ln}(\gamma_1,\gamma_2)$'s are two-row ($km$)-two-column ($ln$) cofactors of the determinant 
(minors of the permanent) constructed out of the overlaps of the localized space orbitals $u({\bf r}, {\bf R}_j)$
[Eq.\ (\ref{uorb})]. The $\mp$ sign in Eq.\ (\ref{cpd2}) corresponds to fermions or bosons.\\
~~~~\\

\section{Single-Particle Density}
\label{appc}

The explicit expression for the SPDs of the broken-symmetry wave packets $\Phi^{\rm PIN}(L_1,L_2;t)$ 
[see Eq.\ (\ref{rho})] is given by
\begin{align}
& \rho({\bf r};t) = \nonumber \\
& \frac{ \int_0^{2\pi} d\gamma_1 \int_0^{2\pi} d\gamma_2 
\left( \alpha^2 e^{i(\gamma_1-\gamma_2)L_1} +
\alpha \beta e^{i(\gamma_1 L_1-\gamma_2 L_2-\phi(t))} +
\alpha \beta e^{i(\gamma_1 L_2+\phi(t)-\gamma_2 L_1)} +
\beta^2 e^{i(\gamma_1-\gamma_2) L_2} \right)
\sum_{kl}{\cal F}_{kl}(\gamma_1,\gamma_2) 
{\cal S}^k_l(\gamma_1,\gamma_2)} {2 \pi \int_0^{2\pi} n(\gamma) 
\left( \alpha^2 e^{i\gamma L_1} + \beta^2 e^{i\gamma L_2} \right)  d\gamma},
\label{rho2}
\tag{A.3}
\end{align}
where 
\begin{align}
{\cal F}_{kl}&(\gamma_1,\gamma_2)= \frac{1}{\pi \lambda^2} \exp \bigg( -\frac
{({\bf r}-{\bf R}_k(\gamma_1))^2 + ({\bf r}-{\bf R}_l(\gamma_2))^2}
{2 \lambda^2} \bigg) \times \nonumber \\
&\exp \bigg( -i \frac
{ y(X_l(\gamma_2)-X_k(\gamma_1))+ x(Y_k(\gamma_1)-Y_l(\gamma_2))}
{2 l_B^2} \bigg),
\label{fkern}
\tag{A.4}
\end{align} 
and the ${\cal S}^k_l(\gamma_1,\gamma_2)$'s are one-row ($k$)-one-column ($l$) cofactors of the determinant 
(minors of the permanent) constructed out of the overlaps of the localized space orbitals $u({\bf r}, {\bf R}_j)$ 
[Eq.\ (\ref{uorb})].\\
~~~~~\\
\end{widetext}


\begin{thebibliography}{99}
\bibitem{bloc08}
I. Bloch, J. Dalibard, and W. Zwerger,
Many-body physics with ultracold gases,
Rev. Mod. Phys. {\bf 80}, 885 (2008).
\bibitem{blat12}
R. Blatt and C.F. Roos,
Quantum simulations with trapped ions,
Nature Phys. {\bf  8}, 277 (2012),
\bibitem{spie14}
N. Goldman, G. Juzeli\~{u}nas, P. \"{O}hberg, and I.B. Spielman,
Light-induced gauge fields for ultracold atoms,
Rep. Prog. Phys. {\bf 77}, 126401 (2014).
\bibitem{taba15}
B. Tabakov, F. Benito, M. Blain, C.R. Clark, S. Clark, R.A. Haltli, P. Maunz, J.D. Sterk, Ch. Tigges, and D. Stick,
Assembling a ring-shaped crystal in a microfabricated surface ion trap,
Phys. Rev. Applied {\bf 4}, 031001 (2015).
\bibitem{haef15}
P.-J. Wang, T. Li, C. Noel, A. Chuang, X. Zhang, and H. H\"{a}ffner,
Surface traps for freely rotating ion ring crystals,
J. Phys. B: At. Mol. Opt. Phys. {\bf 48}, 205002 (2015).
\bibitem{haef17}
H.-K. Li, E. Urban, C. Noel, A. Chuang, Y. Xia, A. Ransford,
B. Hemmerling, Y. Wang, T. Li, H. H\"{a}ffner, and X. Zhang,
Realization of translational symmetry in trapped cold ion rings,
Phys. Rev. Lett. {\bf 118}, 053001 (2017).
\bibitem{nogu14}
A. Noguchi, Y. Shikano, K. Toyoda, and S. Urabe,
Aharonov–Bohm effect in the tunnelling of a quantum rotor in a linear Paul trap,
Nat. Commun. {\bf 5}, 3868 (2014).
\bibitem{bosh14}
C. Ryu, K.C. Henderson, and M.G. Boshier,
Creation of matter wave Bessel beams and observation of quantized circulation in a
BoseEinstein condensate,
New J. Phys. {\bf 16}, 013046 (2014).
\bibitem{moul12}
S. Moulder, S. Beattie, R.P. Smith, N. Tammuz, and Z. Hadzibabic,
Quantized supercurrent decay in an annular Bose-Einstein condensate,
Phys. Rev. A {\bf 86}, 013629 (2012).
\bibitem{phil11}
A. Ramanathan, K.C. Wright, S.R. Muniz, M. Zelan, W.T. Hill, III, C.J. Lobb, K. Helmerson, 
W.D. Phillips, and G.K. Campbell,
Superflow in a toroidal Bose-Einstein condensate: An atom circuit with a tunable weak link,
Phys. Rev. Lett. {\bf 106}, 130401 (2011).
\bibitem{note1}
Refs.\ \cite{taba15,haef15,haef17,bosh14,moul12,phil11} describe specifically recent experimental 
advances in the area of ultracold ring-shaped traps. 
\bibitem{wilc12}
F. Wilczek,
Quantum time crystals,
Phys. Rev. Lett. {\bf 109}, 160401 (2012).
\bibitem{li12}
T. Li, Z-X. Gong, Z-Q. Yin, H.T. Quan, X. Yin, P. Zhang, L-M. Duan, and X. Zhang, 
Space-time crystals of trapped ions, 
Phys. Rev. Lett. {\bf 109}, 163001 (2012).
\bibitem{brun13}
P. Bruno,
Comment on ``Quantum Time Crystals'', 
Phys. Rev. Lett. {\bf 110}, 118901 (2013);
Comment on ``Space-Time Crystals of Trapped Ions'',
Phys. Rev. Lett. {\bf 111}, 029301 (2013).
\bibitem{brun13.2}
P. Bruno,
Impossibility of spontaneously rotating time crystals: A no-go theorem,
Phys. Rev. Lett. {\bf 111}, 070402 (2013).
\bibitem{nozi13}
Ph. Nozi\`{e}res, 
Time crystals: Can diamagnetic currents drive a charge density wave into rotation?,
{\it EPL} {\bf 103}, 57008 (2013).
\bibitem{wilc13}
F. Wilczek,
Wilczek reply,
Phys. Rev. Lett. {\bf 110}, 118902 (2013).
\bibitem{li13}
T. Li, Z.-X. Gong, Z.-Q. Yin, H.T. Quan, X. Yin, P. Zhang, L.-M. Duan, and X. Zhang,
Reply to Comment on ``Space-time crystals of trapped ions,''
arXiv:1212.6959v2. 
\bibitem{wata15}
H. Watanabe and M. Oshikawa,
Absence of quantum time crystals,
Phys. Rev. Lett. {\bf 114}, 251603 (2015).
\bibitem{sach15}
K. Sacha,
Modeling spontaneous breaking of time-translation symmetry, 
Phys. Rev. A {\bf 91}, 033617 (2015). 
\bibitem{sond16}
V. Khemani, A. Lazarides, R. Moessner, and S.L. Sondhi, 
Phase structure of driven quantum systems,
Phys. Rev. Lett. {\bf 116}, 250401 (2016).
\bibitem{naya16}
D.V. Else, B. Bauer, and C. Nayak, 
Floquet time crystals,
Phys. Rev. Lett. {\bf 117}, 090402 (2016).
\bibitem{yao16}
N.Y. Yao, A.C. Potter, I.D. Potirniche, and A. Vishwanath, 
Discrete time crystals: Rigidity, criticality, and realizations,
Phys. Rev. Lett. {\bf 118}, 030401 (2017).
\bibitem{fazi17}
A. Russomanno, F. Iemini, M. Dalmonte, and R. Fazio,
Floquet time crystal in the Lipkin-Meshkov-Glick model,
Phys. Rev. B {\bf 95}, 214307 (2017).
\bibitem{gibn17}
E. Gibney,
The quest to crystallize time,
Nature {\bf 543}, 164 (2017).
\bibitem{monr17}
J. Zhang, P.W. Hess, A. Kyprianidis, P. Becker, A. Lee, J. Smith, G. Pagano, I.-D. Potirniche, A.C. Potter,
A. Vishwanath, N.Y. Yao, and C. Monroe,
Observation of a discrete time crystal,
Nature {\bf 543}, 217 (2017).
\bibitem{luki17}
S. Choi, J. Choi, R. Landig, G. Kucsko, H. Zhou, J. Isoya, F. Jelezko, S. Onoda,
H. Sumiya, V. Khemani, C. von Keyserlingk, N.Y. Yao, E. Demler, and M.D. Lukin,
Observation of discrete time-crystalline order in a disordered dipolar many-body system,
Nature {\bf 543}, 221 (2017).
\bibitem{li12.2}
See also the Supplemental Material of Ref.\ \cite{li12}.
\bibitem{yann07}
C. Yannouleas and U. Landman,
Symmetry breaking and quantum correlations in finite systems: Studies of quantum dots and ultracold Bose 
gases and related nuclear and chemical methods,
Rep. Prog. Phys. {\bf 70}, 2067 (2007). 
\bibitem{degu12}
J. Sato, R. Kanamoto, E. Kaminishi, and T. Deguchi,
Exact relaxation dynamics of a localized many-body state in the 1D Bose gas,
Phys. Rev. Lett. {\bf 108}, 110401 (2012).
\bibitem{kavo17}
G. Eriksson, J. Bengtsson, E.\"{O}. Karabulut, G.M. Kavoulakis, and S.M. Reimann,
Bose-Einstein condensates in a ring potential: Time-evolution beyond the mean-field approximation,
arXiv:1706.00859v1. 
\bibitem{cede11}
A.I. Streltsov, O.E. Alon, and L.S. Cederbaum,
Swift loss of coherence of soliton trains in attractive Bose-Einstein condensates,
Phys. Rev. Lett. {\bf 106}, 240401 (2011).
\bibitem{carr16}
Ch. Weiss and L.D. Carr,
Higher-order quantum bright solitons in Bose-Einstein condensates show truly quantum emergent behavior,
arXiv:1612.05545.
\bibitem{schm17}
G.C. Katsimiga, G.M. Koutentakis, S.I. Mistakidis, P.G. Kevrekidis, and P. Schmelcher,
Dark–bright soliton dynamics beyond the mean-field approximation,
New J. Phys. {\bf 19}, 073004 (2017). 
\bibitem{robi15}
Wave packet dispersion was also reported in a recent study of inhomogeneous spin configurations on a
ring of $N$ bosonic ions, F. Robicheaux and K. Niffenegger,
Quantum simulations of a freely rotating ring of ultracold and identical bosonic ions,
Phys. Rev. A {\bf 91}, 063618 (2015).
\bibitem{yann02}
C. Yannouleas and U. Landman,
Strongly correlated wave functions for artificial atoms and molecules,
J. Phys.: Condens. Matter {\bf 14}, L591 (2002).
\bibitem{yann02.2}
C. Yannouleas and U. Landman,
Trial wave functions with long-range Coulomb correlations for two-dimensional $N$-electron systems 
in high magnetic fields,
Phys. Rev. B {\bf 66}, 115315 (2002).
\bibitem{yann03.2}
C. Yannouleas and U. Landman,
Two-dimensional quantum dots in high magnetic fields:
Rotating-electron-molecule versus composite-fermion approach,
Phys. Rev. B {\bf 68}, 035326 (2003).
\bibitem{yann03}
C. Yannouleas and U. Landman,
Group theoretical analysis of symmetry breaking in two-dimensional quantum dots,
Phys. Rev. B {\bf 68}, 035325 (2003).
\bibitem{yann04}
C. Yannouleas and U. Landman,
Unified description of floppy and rigid rotating Wigner molecules formed in quantum dots,
Phys. Rev. B {\bf 69}, 113306 (2004).
\bibitem{yann04.2}
I. Romanovsky, C. Yannouleas, and U. Landman,
Crystalline boson phases in harmonic traps: Beyond the Gross-Pitaevskii mean field,
Phys. Rev. Lett. {\bf 93}, 230405 (2004).
\bibitem{yann06}
Y. Li, C. Yannouleas, and U. Landman,
From a few to many electrons in quantum dots under strong magnetic fields: Properties of rotating 
electron molecules with multiple rings,
Phys. Rev. B {\bf 73}, 075301 (2006).
\bibitem{yann09}
I. Romanovsky, C. Yannouleas, and U. Landman,
Edge states in graphene quantum dots: Fractional quantum Hall effect analogies and differences at 
zero magnetic field,
Phys. Rev. B {\bf 79}, 075311 (2009).
\bibitem{yann11}
C. Yannouleas and U. Landman,
Unified microscopic approach to the interplay of pinned-Wigner-solid and liquid behavior of the 
lowest Landau-level states in the neighborhood of $\nu =1/3$,
Phys. Rev. B {\bf 84}, 165327 (2011). 
\bibitem{peie57}
R.E. Peierls and J. Yoccoz,
The collective model of nuclear motion,
Proc. Phys. Soc. (London) A {\bf 70}, 381 (1957).
\bibitem{ringbook}
P. Ring and P. Schuck,
{\it The Nuclear Many-Body Problem\/},
(New York, Springer, 1980) ch. 11.
\bibitem{lowe55}
P.-O. L\"{o}wdin,
Quantum theory of many-particle systems: III. Extension of the Hartree-Fock scheme to include degenerate 
systems and correlation effects,
Phys. Rev. {\bf 97}, 1509 (1955).
\bibitem{robl02}
R. Rodr\'{i}guez-Guzm\'{a}n, J.L. Egido, and L.M. Robledo,
Correlations beyond the mean field in Magnesium isotopes: Angular momentum projection 
and configuration mixing,
Nucl. Phys. A {\bf 709}, 201 (2002).
\bibitem{bend03}
M. Bender, P.-H. Heenen, and P.-G. Reinhard,
Self-consistent mean-field models for nuclear structure,
Rev. Mod. Phys. {\bf 75}, 121 (2003).
\bibitem{doba07}
H. Zdu\'{n}czuk, W. Satu\l a, J. Dobaczewski, and M. Kosmulski,
Angular momentum projection of cranked Hartree-Fock states:
Application to terminating bands in $A \sim 44$ nuclei,
Phys. Rev. C {\bf 76}, 044304 (2007).
\bibitem{sun16}
Y. Sun,
Projection-technique approach to the nuclear many-body problem,
Phys. Scr. {\bf 91}, 043005 (2016).
\bibitem{ring00}
J.A. Sheikh and P. Ring,
Symmetry-projected Hartree-Fock-Bogoliubov equations,
Nucl. Phys. A {\bf 665}, 71 (2000).
\bibitem{naza12}
W. Satu\l a, J. Dobaczewski, W. Nazarewicz, and T.R. Werner
Isospin-breaking corrections to superallowed Fermi $\beta$ decay in isospin- and
angular-momentum-projected nuclear density functional theory,
Phys. Rev. C {\bf 86}, 054316 (2012).
\bibitem{giulbook}
G. Giuliani and G. Vignale,
{\it Quantum Theory of the Electron Liquid\/} 
(Cambridge University Press, Cambridge, 2008) ch 1.6
\bibitem{grif76}
P.C. Lichtner and J.J. Griffin,
Evolution of a quantum qystem: Lifetime of a determinant,
Phys. Rev. Lett. {\bf 37}, 1521 (1976).
\bibitem{maki83}
\textcolor{black}{
K. Maki and X. Zotos,
Static and dynamic properties of a two-dimensional Wigner crystal in a strong magnetic field,
Phys. Rev. B {\bf 28}, 4349 (1983).}
\bibitem{landbook}
L.D. Landau and E.M. Lifshitz,
{\it Course of Theoretical Physics\/}, Vol. 9,
Statistical Physics, Part 2
(Pergamon Press, Oxford, 1980) p. 248.
\bibitem{peie33}
R.E. Peierls,
Zur theorie des diamagnetismus von leitungselektronen,
Z. Physik {\bf 80}, 763 (1933).
\bibitem{wolbbook}
A.B. Wolbarst, 
{\it Symmetry and Quantum Systems\/} 
(Van Nostrand Reinold, New York, 1977).
\bibitem{cottbook}
F.A. Cotton, 
{\it Chemical Applications of Group Theory\/},
(Wiley, New York, 1990).
\bibitem{hamebook}
M. Hamermesh,
{\it Group Theory and its Application to Physical Problems\/}
(Addison-Wesley, Reading, MA, 1962).
\bibitem{hurlbook}
A.C. Hurley,
{\it Introduction to the Electron Theory of Small Molecules\/}
(Academic Press, London, 1976) ch. 6.
\bibitem{ishi03}
K. Ishida,
Molecular integrals over the gauge-including atomic orbitals,
J. Chem. Phys. {\bf 118}, 4819 (2003).
\bibitem{boys50}
S.F. Boys,
Electronic wave functions. I. A general method of calculation for the stationary states of 
any molecular system, 
Proc. R. Soc. (London) Ser. A {\bf 200}, 542 (1950).
\bibitem{yannpub}
C. Yannouleas and U. Landman,
to be published.
\bibitem{yann00}
\textcolor{black}{
C. Yannouleas and U. Landman,
Collective and independent-particle motion in two-electron artificial atoms,
Phys. Rev. Lett. {\bf 85}, 1726 (2000).}
\bibitem{kana08}
\textcolor{black}{
R. Kanamoto, L.D. Carr, and M. Ueda,
Topological winding and unwinding in metastable Bose-Einstein condensates,
Phys. Rev. Lett. {\bf 100}, 060401 (2008).}
\bibitem{kana09}
\textcolor{black}{
R. Kanamoto, L.D. Carr, and M. Ueda,
Metastable quantum phase transitions in a periodic one-dimensional Bose gas: Mean-field and Bogoliubov analyses,
Phys. Rev. A {\bf 79}, 063616 (2009).}
\bibitem{kana10}
\textcolor{black}{
R. Kanamoto, L.D. Carr, and M. Ueda,
Metastable quantum phase transitions in a periodic one-dimensional Bose gas. II. Many-body theory,
Phys. Rev. A {\bf 81}, 023625 (2010).}
\bibitem{bomo98}
\textcolor{black}
{\AA. Bohr \& B.R. Mottelson, 
{\it Nuclear Structure\/}, 
(World Scientific, Singapore, 1998) Vol. II.}
\bibitem{ruan95}
\textcolor{black}{
W.Y. Ruan, Y.Y. Liu, C.G. Bao, and Z.Q. Zhang,
Origin of magic angular momenta in few-electron quantum dots,
Phys. Rev. B {\bf 51}, 7942 (1995).}
\bibitem{maks96}
\textcolor{black}{
P.A. Maksym,
Eckardt frame theory of interacting electrons in quantum dots,
Phys. Rev. B {\bf 53}, 10871 (1996).}
\bibitem{seki96}
\textcolor{black}{
T. Seki, Y. Kuramoto, and T. Nishino,
Origin of magic angular momentum in a quantum dot under strong magnetic field,
J. Phys. Soc. Japan. {\bf 65}, 3945 (1996).}
\bibitem{maks00}
\textcolor{black}{
P.A. Maksym, H. Imamura, G.P. Mallon, and H. Aoki,
Molecular aspects of electron correlation in quantum dots,
J. Phys.: Condens. Matter {\bf 12}, R299 (2000).}
\bibitem{bocc57}
P. Bocchieri and A. Loinger,
Quantum recurrence theorem,
Phys. Rev. {\bf 107}, 337 (1957).
\bibitem{schu78}
L. S. Schulman,
Note on the quantum recurrence theorem,
Phys. Rev. A {\bf 18} 2379 (1978).
\bibitem{che17}
H. Che, K. Deng, Z.T. Xu, W.H. Yuan, J. Zhang, and Z.H. Lu,
Efficient Raman sideband cooling of trapped ions to their motional ground state,
Phys. Rev. A {\bf 96}, 013417 (2017).
\end{thebibliography}
\end{document}